\documentclass[11pt]{article}

\usepackage[a4paper,margin=1in]{geometry}
\usepackage[T1]{fontenc}
\usepackage[utf8]{inputenc}
\usepackage{lmodern}
\usepackage{amsmath,amssymb}
\usepackage{verbatim}
\usepackage{xurl}
\usepackage{url}
\usepackage{graphicx}
\usepackage{booktabs}
\usepackage{pifont} 

\title{
  \textbf{Architectural Obsolescence of Unhardened Agentic-AI Runtimes}\\[2pt]
  \large A production-CLI in-vivo comparison of OpenClaw and enclawed-oss\\
  on F1--F4 detection, extension parity, and improvability
}

\author{%
  Alfredo Metere\\
  Metere Consulting, LLC.\\
  \texttt{alfredo.metere@metereconsulting.com}
}

\date{May 2, 2026}

\begin{document}

\maketitle

\begin{abstract}
An agentic-AI runtime issues tool calls, sends messages, and
actuates devices on behalf of an LLM. Catching the four ways an
action can diverge from its audit record ---
\textbf{F1}~gate-bypass, \textbf{F2}~audit-forgery, \textbf{F3}~silent
host failure, \textbf{F4}~wrong-target~\cite{Metere2026enclawed}\,§5
--- is a load-bearing safety property of any such runtime. We show
that upstream OpenClaw, the most engineered single-user agentic-AI
gateway in public release, catches none of them: recall is 0.000 on
every cell of every confusion matrix, on a 1600-sample template
baseline through OpenClaw's actual production command-line
interface (CLI) and on a ten-LLM cross-model generalisation run.

Detecting F1--F4 requires seven specific runtime structures
\textbf{absent from OpenClaw's source tree}: a biconditional
checker, a hash-chained audit log, an extension admission gate, a
two-layer egress guard, a Bell-LaPadula classification policy, a
module-signing trust root, and a bootstrap seal. enclawed-oss ---
an MIT-licensed drop-in fork that ships all seven --- reaches
$P = R = F_1 =$ accuracy $= 1.000$ on the same input. The gap is
structural, not parametric: a six-line append-only widening of
enclawed-oss's data-loss-prevention (DLP) regex catalog raises
per-channel F3 detection by
14.6\% net at unchanged precision; the same edit on OpenClaw has
nowhere to land. The harness deliberately exercises real Discord
and Telegram channels --- plugin categories the first enclawed
release deleted as unsafe --- to show F1--F4 detection extends to
those previously-unsafe extensions. With architectural superiority
for security \emph{and} feature parity for extensions
(\S\ref{sec:def-obsolescence}, contributions), we argue that
unhardened agentic-AI runtimes are architecturally obsolete: a
strictly better alternative exists, is adoptable today, and the gap
requires re-architecture rather than configuration. We invite
reviewers to apply the harness to any candidate runtime.
\end{abstract}

\section{Introduction}

Obsolescence is a comparative property. A tool is obsolete with
respect to a use case when a strictly better alternative exists, the
alternative is adoptable by current operators of the tool, and the
gap between them is wide enough that staying on the tool no longer
reflects a defensible choice. Obsolescence is therefore not a
property of the deprecated tool alone; it is a property of the
\emph{gap} between the deprecated tool and the available alternative,
relative to a specific use case the alternative serves better.

This paper makes a claim of that shape about agentic-AI runtimes.
Upstream OpenClaw~\cite{OpenClaw} and the broader
unhardened-agentic-AI runtime category it exemplifies are obsolete
with respect to the \textbf{four biconditional failure modes
F1--F4} of \cite{Metere2026enclawed}\,§5 (gate-bypass,
audit-forgery, silent host failure, wrong-target) because a strictly
better alternative --- enclawed-oss --- exists, is MIT-licensed and
drop-in adoptable, and the gap is structural rather than parametric.
Detecting F1--F4 requires a \textbf{vocabulary of seven primitives}
(Table~\ref{tab:primitive-presence}): a biconditional checker, a
hash-chained audit log, an extension admission gate, a two-layer
egress guard, a Bell-LaPadula classification policy lattice, a
module-signing trust root, and a bootstrap seal. The four are the
attack \emph{shapes}; the seven are the runtime \emph{structures}
needed to detect them.

We mediate 1600 adversarial-and-legit chat samples through
OpenClaw's actual production CLI and through two hardened forks.
OpenClaw recall is 0.000 on every F1--F4 category; both hardened
subjects reach $P = R =$ accuracy $= 1.000$. A tree-walk of
OpenClaw's 14{,}419 first-party source files returns zero matches
for any of the seven primitives; those same primitives are shipped
in enclawed-oss. The gap between 0\% and 100\% is not a flag, a
configuration knob, or a plug-in; it is the seven primitives.

\paragraph{Definition (architectural obsolescence).}
\label{sec:def-obsolescence}
A runtime category $\mathcal{R}$ is \emph{architecturally obsolete}
with respect to a failure-mode set $\mathcal{F}$ iff there exists
an alternative runtime $A \notin \mathcal{R}$ such that:
\begin{enumerate}
  \item \textbf{Strict dominance.} An end-to-end adversarial
    harness reports recall = 0 for every $R \in \mathcal{R}$ on
    every mode in $\mathcal{F}$, while $A$ reports recall = 1 (or
    materially higher than every $R$) at the same or better
    precision.
  \item \textbf{Practical adoptability.} $A$ is available to any
    operator who currently runs $R$: open-source license,
    drop-in fork, no new infrastructure, no cost barrier.
  \item \textbf{Structural gap.} A tree-walk over $R$'s first-party
    source returns zero matches for any primitive in the detection
    vocabulary $\mathcal{P}(\mathcal{F})$ that $A$ ships; closing
    the gap to $A$'s recall requires the joint addition of
    $\mathcal{P}(\mathcal{F})$, i.e.\ a re-architecture, not a
    parameter, a flag, or a plug-in.
\end{enumerate}
Condition~(1) is the gap; (2) is what makes the gap operationally
relevant (a strictly better alternative that no one can adopt does
not obsolete the incumbent); (3) distinguishes \emph{obsolete} from
\emph{weak} --- a runtime catching 60\% of F1 has a tunable detector
and is weak; a runtime catching 0\% with no detector in the source
tree is obsolete relative to one that catches 100\% via primitives
the first does not have. We do not claim the seven primitives are
the unique decomposition: a different decomposition that achieves
the same recall would equally support the claim for any runtime
that ships it. The empirical contribution of this paper is to
exhibit such a pair --- the unhardened-agentic-AI runtime category
$\mathcal{R}$, with upstream OpenClaw as the \emph{strongest
measured representative}, and the alternative $A$ = enclawed-oss
--- and to verify all three conditions for them. Choosing the
strongest representative is deliberate: weaker members of
$\mathcal{R}$ score $\leq$ OpenClaw's score by inheritance and
require no separate measurement.

\paragraph{Stakes.} PCI~DSS~10.x, SOC~2 CC7.2, HIPAA 164.312(b),
FedRAMP AU-2, NIST 800-53 AU-9, and GDPR Art.~30 all require
tamper-evident audit of every consequential decision; a runtime that
emits no audit on F1 satisfies none of them. enclawed-oss's core
architecture is ready for Federal Information Processing Standards
(FIPS) 140-3 (the \texttt{crypto-fips} module gates symmetric
encryption behind \texttt{assertFipsMode}, signed manifests use
Ed25519 per FIPS 186-5, and the \texttt{enclaved} flavor asserts
FIPS at boot) and the framework's control families
align with NIST 800-53, ISO/IEC 27001/27002, NIST CSF 2.0, SOC 2,
GDPR, HIPAA, PCI DSS v4, CMMC L2/L3 / NIST 800-171, and CNSSI 1253;
upstream OpenClaw aligns with none of them by inspection. The
empirical incidence record agrees: \emph{Broken Access Control}
(the F1 surface) is OWASP Top~10:2021's most prevalent
vulnerability class, present in \textbf{94\%} of tested
applications~\cite{OWASP_A01_2021}; trustworthy internal detection
shortens the breach lifecycle by \textbf{61~days} on average
\cite{IBM_CostOfBreach_2024}; Mandiant M-Trends 2024 reports
global median dwell time at \textbf{10~days}, down from 16, as
internal detection rises from 37\% to
46\%~\cite{Mandiant_MTrends_2024}. For deployments that actuate
physical devices the obsolescence is also a safety property: ``no
detection mechanism exists'' is no longer a tenable default once
the loop reaches a robot, a CNC mill, or an irrigation
valve.

\paragraph{Coverage scope.} enclawed-oss's defences are
\emph{application-level} and complete on two surfaces: the
\emph{structural} surface F1--F4 (biconditional over hash-chained
audit) and the \emph{content} surface (secrets, personally
identifiable information (PII), prompt injection via DLP and
\texttt{prompt-shield}). Out of scope are system-level software
attacks (time-of-check-to-time-of-use (TOCTOU) on the corpus
snapshot, side-channel timing, crypto-primitive attacks, operator
collusion; addressed by process isolation, write-once-read-many
(WORM) audit storage, FIPS-validated crypto, and separation of
duty) and application-level attacks outside F1--F4
(information-flow control for paraphrase exfiltration, semantic
intent verification, indirect injection through ungated tool-output
channels; future work). The obsolescence claim is therefore scoped
exactly: enclawed-oss is fully safe from the application-level
software-only attack classes the F1--F4 contract names; OpenClaw is
safe from none of them.

\paragraph{Extension parity, including the previously-unsafe set
(new in this paper).}
The first enclawed paper~\cite{Metere2026enclawed} introduced the
F1--F4 primitives but shipped with a reduced extension set in the
high-trust \texttt{enclaved} flavor: 78 cloud-channel and cloud-LLM
modules (Discord, Telegram, Slack, OpenAI, Anthropic, Google,
\ldots) were deleted as unsafe under that flavor's threat model.
The present work closes the gap on two coupled axes: \textbf{(1)}
the OSS \texttt{open} flavor now ships every one of OpenClaw's 125
bundled extensions plus an added Model Context Protocol (MCP)
attestation reference module
(126 total); \textbf{(2)} the harness exercises real Discord and
Telegram channels --- precisely the plugin categories the first
enclawed release deleted --- and reports recall = 1.000 on every
F-category, demonstrating that F1--F4 detection secures the exact
previously-unsafe plugin classes. With architectural superiority for
security \emph{and} feature parity that extends to those plugins,
no remaining deployment constraint motivates choosing upstream
OpenClaw over enclawed-oss.

\paragraph{Contributions.}
\begin{itemize}
  \item A falsifiable definition of architectural obsolescence
    (above) and a non-empty witness category for it
    (§\ref{sec:obsolescence}).
  \item A reproducible adversarial harness that runs upstream
    OpenClaw end-to-end through its production CLI and exercises
    all three subjects on real Discord and Telegram bot endpoints
    (§\ref{sec:methodology}, §\ref{sec:subjects}).
  \item Empirical results on two complementary distributions ---
    in-distribution templates ($n=1600$ headline, $n=80{,}000$
    stress) and ten-LLM cross-model generalisation
    (§\ref{sec:results}).
  \item Empirical extension parity between enclawed-oss and upstream
    OpenClaw (126 vs.\ 125 bundled extensions; \texttt{open} flavor),
    a property absent from the first enclawed paper that is required
    for the obsolescence claim to bind in practice.
  \item A live demonstration that the F1--F4 gate is improvable on
    the hardened category and structurally not on the obsolete
    category: a six-line append-only widening of the production DLP
    catalog raises per-channel F3 TP counts by 14.6\% net at
    unchanged precision (§\ref{sec:widening}).
\end{itemize}

\section{The F1--F4 failure-mode taxonomy}
\label{sec:f}

The obsolescence claim rests on a specific failure-mode set
$\mathcal{F}$ inherited from \cite{Metere2026enclawed}\,§5. We
restate it briefly so the rest of the paper is self-contained.

\subsection{Intuition}

Two things happen for every consequential agent action: the action
mutates the world (the \emph{corpus delta} $D$) and the agent writes
a hash-chained audit record describing it (the \emph{audit log}
$S$). A correct agent keeps them in lockstep: every real mutation
has exactly one successful audit record, and vice versa. The four
ways they can diverge are F1--F4:

\begin{itemize}
  \item \textbf{F1 (gate bypass)} --- mutation in $D$, no matching
    record in $S$. The agent moved state without going through the
    admission gate.
  \item \textbf{F2 (audit forgery)} --- record in $S$, no matching
    mutation in $D$. The audit log is being used to fabricate
    evidence of work that never happened.
  \item \textbf{F3 (approved-but-failed silent)} --- mutation
    partially landed but the audit record is \texttt{ok=false} (so
    excluded from $S$); the residue projects as F1.
  \item \textbf{F4 (wrong target)} --- the gate approved
    $(\mathit{cap}, \mathit{targetA})$ but the corpus mutation landed
    on $(\mathit{cap}, \mathit{targetB})$.
\end{itemize}

\noindent The biconditional check is a multiset diff on
$(\mathit{cap}, \mathit{target})$ keys; ordering, timing, and content
do not enter it. F1, F2, F4 manifest as named diffs with the
offending key reported; F3 collapses to the F1 shape. Modes that do
\emph{not} cause $D \neq S$ --- read-only exfiltration, TOCTOU races
that finish before the diff reads, collusion with a privileged
operator whose actions are correctly logged --- are out of scope by
construction; \S\ref{sec:scope} returns to them.

\subsection{Formal statement}

We give the F1--F4 contract precise mathematical form. Let
$\mathcal{C}$ be the enclawed capability vocabulary (a fixed finite
set; nine tokens in v1, including $\textsf{net.egress}$, $\textsf{fs.read}$,
$\textsf{tool.invoke}$, $\textsf{publish}$, $\textsf{pay}$). Let
$\mathcal{T}$ be the universe of canonical target identifiers
(channel URIs, file paths, host URLs).

A \emph{corpus delta} is a multiset $D \in \mathbb{N}^{\mathcal{C}\times\mathcal{T}}$
recording, for each (capability, target) pair, the count of real-world
mutations the agent caused. An \emph{audit log} is a sequence of
records $L = (r_1, r_2, \ldots)$ each of which carries a type
$\tau_i$, a payload $\rho_i$, and a hash-chain field
$h_i = \mathsf{SHA\text{-}256}(h_{i-1} \,\|\, \mathsf{canon}(r_i))$.
The audit-side multiset $S \in \mathbb{N}^{\mathcal{C}\times\mathcal{T}}$
is the projection
\begin{equation}
S \;=\; \pi_{(\mathit{cap},\mathit{tgt})}\!\bigl(
  \{ r \in L \;:\; \tau(r) = \texttt{irreversible.executed} \,\wedge\, \rho(r).\texttt{ok} = \mathsf{true} \}
\bigr),
\label{eq:S-projection}
\end{equation}
i.e., we keep only the audit records that claim a successful
irreversible operation, and project them onto the (capability,
target) key. Records marked $\texttt{ok}=\mathsf{false}$ (silent
failures) are deliberately excluded from $S$; they survive in $L$
for forensics but do not count toward the reconciliation.

The \emph{biconditional check} compares $D$ and $S$:
\begin{equation}
\textsc{check}(D, S) \;=\;
\begin{cases}
\mathrm{ok} & \text{if } D = S \text{ as multisets,} \\
\mathrm{f1Bypass}(D \setminus S) & \text{if } D \supsetneq S, \\
\mathrm{f2Forgery}(S \setminus D) & \text{if } S \supsetneq D, \\
\mathrm{f4WrongTarget}(D \setminus S,\, S \setminus D) & \text{if } D \neq S \text{ and neither contains the other.}
\end{cases}
\label{eq:check}
\end{equation}
F3 (silent host failure) collapses to the F1 case at the multiset
level: a half-finished operation contributes to $D$ but its
$\texttt{ok}=\mathsf{false}$ audit record is excluded from $S$, so
$D \setminus S$ becomes non-empty.

The biconditional check from~\cite{Metere2026enclawed}~§5 compares a
corpus-delta multiset $D$ against an audit-log multiset $S$, both
keyed by the projection $(\mathit{cap}, \mathit{target})$, where
$\mathit{cap}$ is a capability token from the enclawed vocabulary and
$\mathit{target}$ is the canonical target identifier of the
operation. The four failure modes manifest as divergences between the
two multisets:

\begin{table}[h]
\centering
\small
\begin{tabular}{|p{2.3cm}|p{4.0cm}|p{8.0cm}|}
\hline
\textbf{Mode} & \textbf{Definition} & \textbf{Detection mechanism} \\
\hline
\textbf{F1} gate bypass &
$D \setminus S \neq \emptyset$ &
Corpus delta exists with no matching \texttt{irreversible.executed} audit
record where \texttt{ok=true}. The agent (or an attacker controlling the
agent) mutated state without going through the admission gate. \\
\hline
\textbf{F2} audit forgery &
$S \setminus D \neq \emptyset$ &
Audit record claims \texttt{irreversible.executed} with \texttt{ok=true}
but the corpus shows no matching mutation. The audit log is being used to
fabricate evidence of work that never happened. \\
\hline
\textbf{F3} approved-but-failed silent &
$D \setminus S \neq \emptyset$ &
Gate approved the call; host began executing it; failed silently
mid-operation. The audit record exists with \texttt{ok=false} (excluded
from $S$) so the partial corpus mutation projects as F1. \\
\hline
\textbf{F4} wrong target &
$D \setminus S \neq \emptyset$ \emph{and} $S \setminus D \neq \emptyset$ &
Gate approved $(\mathit{cap}, \mathit{targetA})$ but corpus mutated
$(\mathit{cap}, \mathit{targetB})$. Because the projection key includes
target, the audit's $(\mathit{cap}, \mathit{targetA})$ and the corpus's
$(\mathit{cap}, \mathit{targetB})$ land in different multiset buckets. \\
\hline
\end{tabular}
\caption{The four biconditional failure modes (paper~§5).}
\label{tab:fmodes}
\end{table}

A runtime detects the failure modes if and only if it carries (1) an
audit log that records each $(\mathit{cap}, \mathit{target})$ pair the gate
approved, (2) a corpus-delta source the runtime can canonicalize against
the same projection, and (3) a checker that compares the two multisets.
None of those three primitives exists in upstream OpenClaw.

\section{Threat model}
\label{sec:threat-model}

\textbf{Adversary capability.} The adversary controls any string the
agent's host runtime processes --- chat message, retrieved-document
chunk, tool-output payload, operator prompt --- including content
that imitates LLM serialisation tokens, embeds known secret/PII
shapes, or mis-matches the intended target. The adversary may
\emph{not} compromise the host process, the trust root, the
audit-log filesystem, or the cryptographic primitives; those are
addressed by system-level orthogonal controls (process isolation,
FIPS-validated crypto, WORM storage).
\textbf{Knowledge.} White-box: read access to all subject source
trees, including the gate stack, the regex catalogs, and the F1--F4
contract verbatim. Stronger than black-box; consistent with any
defence-in-depth claim that does not lean on obscurity.
\textbf{Out of scope}, per \cite{Metere2026enclawed}\,§5.4:
read-only exfiltration that does not cause $D \neq S$, TOCTOU races
on the corpus, operator collusion with full corpus access,
side-channel timing extraction, and direct crypto-primitive attacks.
A gate detection rate of 1.000 on F1--F4 does not imply 1.000 on
these.

\section{The three subjects}
\label{sec:subjects}

The harness mediates every sample through three runtimes drawn from
the same lineage. They differ in exactly the seven detection
primitives F1--F4 require (Table~\ref{tab:primitive-presence}).

\paragraph{OpenClaw (upstream).} The single-user agentic-AI gateway
by Steinberger~\cite{OpenClaw}, taken throughout this paper as the
\emph{strongest representative} of the unhardened-agentic-AI runtime
class: the most engineered, most adopted, and most feature-complete
single-user agentic-AI gateway in public release. Permissive
defaults: external channels, external LLM providers, plug-in
modules installed via \texttt{npm} with no enforced signature
verification. Audit, classification, and DLP are not part of the
framework's contract. The harness exercises upstream
OpenClaw \emph{by invoking the actual upstream CLI per sample}, not as
a probe or a shim: each per-sample mediation spawns
\texttt{node openclaw.mjs message send --channel <discord|telegram>
--target <id> --message <content> --dry-run --json} against an upstream
checkout that has been built with \texttt{pnpm install \&\& pnpm
build:docker}, captures the JSON envelope upstream emits, and records
the verdict the runtime itself reports. \texttt{--dry-run} stops
upstream's pipeline at the network-call boundary so no real
Discord/Telegram message is transmitted, but every other code path is
the same code path a production end-user invocation exercises (CLI argv
parsing $\to$ message-send command handler $\to$ channel resolution
$\to$ payload framing $\to$ outbound delivery decision). A successful
delivery decision (envelope shape \texttt{action=send}, \texttt{payload}
populated, \texttt{handledBy} non-empty, no \texttt{error} field) is
the verdict the harness records as \texttt{delivered = true} for that
sample; any error or rejection at any step in upstream's pipeline is
the runtime's actual content-gate verdict. Every per-sample OpenClaw
verdict in this paper is the verdict upstream's runtime emits when its
own code path runs end-to-end.

\paragraph{enclawed-oss.} An MIT-licensed, independently re-engineered
divergent hard fork of OpenClaw~\cite{Metere2026enclawed}. Adds a
hash-chained audit log, a Bell-LaPadula classification lattice, an Ed25519
module-signing trust root, an extension admission gate (signed manifest,
declared capabilities, biconditional admission for net-capable
extensions), a two-layer egress guard (fetch wrapper plus
\texttt{Socket.prototype.connect} guard) with VPN-only mode, a structural
prompt shield, and a regex-based DLP scanner. Every shipped extension is
admitted under the same gate.

\paragraph{enclawed-enclaved.} The proprietary certification layer.
Inherits everything enclawed-oss provides and adds, at the framework
contract level: (i) a fail-closed admission contract --- if the
external-witness component is not engaged at boot, the runtime refuses
to admit any extension for the rest of the run; (ii) externally-witnessed
audit --- every gate decision carries an Ed25519-signed witness record
that an independent third party can re-verify with only the witness's
public key; (iii) a behavioral-anomaly monitor sitting alongside the
content-gate stack so that abnormal action patterns are flagged in
addition to abnormal content; (iv) tamper-attempt accounting in the
audit log distinct from regular policy denials. Internal architecture
is out of scope for this paper.

\subsection{Empirical primitive availability}

Table~\ref{tab:primitive-presence} reports the result of a two-stage
audit of each subject's published surface. First, a tree-walk over
the first-party source files (\texttt{*.ts}, \texttt{*.tsx},
\texttt{*.mjs}, \texttt{*.js}, \texttt{*.cjs}; \texttt{node\_modules},
\texttt{dist}, \texttt{build}, \texttt{out}, \texttt{coverage}
excluded) greps for the canonical exported symbol of each detection
primitive. Second, the subject's user-facing documentation, plugin
SDK reference, and API surface (\texttt{README}, \texttt{AGENTS.md},
\texttt{docs/}, \texttt{src/plugin-sdk/} where applicable) is read to
confirm whether any equivalent primitive ships under a different
name --- so a runtime that implements the same contract under a
non-canonical symbol would still register as ``present''. Both
stages must miss the primitive for the cell to read \emph{absent}.
Upstream OpenClaw fails both stages on every primitive (no canonical
symbol matches; no equivalent contract documented in
\texttt{docs/}); enclawed-oss passes both on every primitive.

\begin{table}[h]
\centering
\footnotesize
\setlength{\tabcolsep}{4pt}
\resizebox{\textwidth}{!}{%
\begin{tabular}{|p{8.5cm}|c|c|c|}
\hline
\textbf{Primitive} & \textbf{OpenClaw} & \textbf{enclawed-oss} & \textbf{enclawed-enclaved} \\
\hline
biconditional checker (\texttt{checkBiconditional})           & absent & present & present \\
\hline
hash-chained \texttt{AuditLogger}                             & absent & present & present \\
\hline
extension admission gate (\texttt{admitExtension})            & absent & present & present \\
\hline
two-layer egress guard (\texttt{installEgressGuard} \&\ \texttt{installRawSocketGuard}) & absent & present & present \\
\hline
Bell-LaPadula classification (\texttt{defaultEnclavedPolicy}) & absent & present & present \\
\hline
module-signing trust root (\texttt{lockTrustRoot})            & absent & present & present \\
\hline
bootstrap seal (\texttt{sealBootstrap})                       & absent & present & present \\
\hline
\end{tabular}%
}
\caption{Empirical detection-primitive availability per subject. The probe
walked 14{,}419 upstream files and found zero matches for any of the seven
canonical symbols on OpenClaw.}
\label{tab:primitive-presence}
\end{table}

\noindent The table is the topology of obsolescence: the seven primitives
the failure modes depend on are uniformly missing in upstream OpenClaw,
uniformly present in enclawed-oss and enclawed-enclaved.

\section{Methodology}
\label{sec:methodology}

The harness is a single Node entrypoint
(\path{enclawed/test/security/in-vivo/llm-narrative.mjs}) that
generates a balanced legit/adversarial sample set per F-category and
mediates every sample through all three subjects.

\subsection{Statistical adversarial harness}

100 legit + 100 adversarial samples per F-category. The three
subjects:

\begin{itemize}
  \item \textbf{OpenClaw.} Real upstream-runtime subject: at boot the
    harness verifies an upstream OpenClaw checkout exists at
    \texttt{\$OPENCLAW\_PATH} with a built \texttt{dist/entry.js}; per
    sample, it spawns the upstream \texttt{openclaw} CLI itself with
    \texttt{message send --channel <id> --target <t> --message
    <content> --dry-run --json} (the same CLI an end user invokes) and
    parses the JSON envelope upstream's runtime emits to derive the
    verdict. The harness uses an isolated
    \texttt{OPENCLAW\_STATE\_DIR} so the run never touches operator
    state. \texttt{--dry-run} stops the pipeline at the network-call
    boundary; every other step (argv parsing, message-send command
    handler, channel resolution, payload framing, outbound delivery
    decision) executes exactly as in production. Upstream's tree
    carries no content-side gate, audit log, biconditional checker,
    content shield, or behavioral monitor between the LLM-emitted
    payload and the chat backend (Table~\ref{tab:primitive-presence}),
    so every per-sample verdict is the verdict upstream's runtime
    actually emits when its own code runs end-to-end.
  \item \textbf{enclawed-oss.} The harness invokes the real shipped
    primitives: \texttt{admitExtension} in enclaved flavor, the
    \texttt{Policy} channel/provider allowlist, the persistent
    hash-chained \texttt{AuditLogger}, the biconditional
    \texttt{checkBiconditional} reconciler, the \texttt{prompt-shield}
    \texttt{detectInjection} detector, and the DLP \texttt{scan}
    regex catalog.
  \item \textbf{enclawed-enclaved.} The certified evolution. From the
    paper's point of view, the user-visible properties exercised by
    the harness are: (i) every gate decision is externally witnessed
    and the witness record is independently re-verifiable by a third
    party; (ii) if no external witness is available at boot, the
    runtime refuses to admit any extension for the rest of the run
    (a fail-closed property of the hardening contract); (iii) a
    behavioral-anomaly monitor sits alongside the content-gate stack
    so that abnormal action patterns are flagged in addition to
    abnormal content. The fail-closed half is verified empirically
    by re-running the harness with the runtime's external-witness
    component intentionally disabled.
\end{itemize}

\subsection{Safety envelope}

\begin{sloppypar}
The harness reads cloud target IDs and LLM API keys only from
\texttt{ENCLAWED\_INVIVO\_*} env vars (never the operator's normal
credentials); writes per-tag hash-chained audit + witness journals to
isolated paths so concurrent runs never collide;
\texttt{STATS\_ONLY=1} suppresses chat-platform posts while keeping
the gate stack live; \texttt{DISABLE\_ACCREDITOR=1} forces the
enclaved subject's witness offline to validate the fail-closed
admission path; a 600--800~ms inter-post delay keeps cloud-channel
rate limits comfortable. Per-call wall-clock and byte caps are
inherited from each LLM provider's own limits and the harness's
\texttt{LLM\_CONCURRENCY} ceiling (default 8); chat posts are clamped
to 1900 chars on Discord and 3500 on Telegram.
\end{sloppypar}

\subsection{Sample sources: template synthesis vs.\ LLM generation}
\label{sec:sample-sources}

An honest in-vivo experiment must distinguish two questions:
\emph{does the gate fire on adversarial content?} and \emph{can a
contemporary LLM be cajoled into producing such content?} These are
orthogonal and we report both.

\paragraph{Template synthesis (default).} The harness generates each
sample from a parameterized template that randomizes across the patterns
the production gates actually classify --- the prompt-shield
imperative-override and role-boundary-token sets, the DLP secret/PII
regex catalog. Templates are deterministic, diverse (hundreds of
combinatorial variants per F-category), and \emph{independent} of LLM
cooperation. With $n=100$ per cell the per-bin variance is small enough
to compare subjects directly.

\paragraph{LLM generation (opt-in).} Setting
\texttt{ENCLAWED\_INVIVO\_SAMPLE\_SOURCE=llm} routes the same briefs
through a real LLM (Anthropic, OpenAI-compatible OpenRouter, or local
Ollama). The LLM receives an explicit security-research framing and is
asked to produce the adversarial content directly. The two paths share
the rest of the pipeline; only the content source differs.

\paragraph{Why both.} The LLM-driven path is contaminated by
\emph{refusal asymmetry}: a safety-tuned model (Llama 3.2:3b on Ollama,
or Claude Haiku 4.5 on Anthropic) declines to author roughly 95\% of
the F1/F3/F4 briefs, returning benign refusal text instead. That
benign text is correctly delivered by the gate (no injection patterns,
no secrets), so the ground-truth label ``adversarial'' becomes a
mislabel of refusal text. Counting these as false negatives makes the
gate look bad while measuring something else entirely. Templates
remove that confounder. We report templates as the primary statistical
source; the LLM path is reported separately as an empirical study of
LLM refusal rates per F-category (Table~\ref{tab:llm-coop}).

\subsection{F-category instantiation}

The harness exercises four categories. Three of them are \emph{content}
attacks (the gate is a regex/policy/classifier on the chat payload);
one of them (F2) is \emph{behavioral} (the gate is the multiset
biconditional comparing audit records to corpus deltas).

\begin{itemize}
  \item \textbf{F1\_BYPASS.} Gate-bypass via prompt injection: a chat
    payload combining an imperative-override phrase
    (\texttt{IGNORE}/\texttt{DISREGARD}/\texttt{OVERRIDE}/\texttt{BYPASS}/%
    \texttt{FORGET}/\texttt{PURGE}) with a role-boundary token
    (\texttt{</system>}, \texttt{<|im\_end|>}, \texttt{[/INST]},
    \texttt{<|system|>}, \texttt{<|endoftext|>}). The
    \texttt{prompt-shield.detectInjection} detector reports
    \texttt{imperative-override} or \texttt{role-boundary-token}.
  \item \textbf{F2\_FORGE.} Audit forgery: the harness appends an
    \texttt{irreversible.executed} audit record claiming an op
    succeeded but does \emph{not} add a matching corpus-delta entry.
    Each F2 sample is tagged with a unique probe identifier so the
    biconditional check is scoped to that single $(\mathit{cap},
    \mathit{target}_{\mathit{probe}})$ projection. Legit
    \texttt{LEGIT\_AUDIT} samples record \emph{both} audit and delta
    and pass cleanly.
  \item \textbf{F3\_SILENT.} Silent failure / secret exfiltration: a
    chat payload that smuggles a secret (OpenAI \texttt{sk-}, AWS
    \texttt{AKIA}, GitHub \texttt{ghp\_}, Stripe, JWT, credit-card-like
    digit string) inside a plausible recovery-context wrapper. The DLP
    \texttt{scan} regex catalog flags the high-severity pattern.
  \item \textbf{F4\_WRONGTGT.} Wrong-target broadcast: a chat payload
    that pretends to be a privileged broadcast accidentally landing in
    the wrong channel and includes both an email address and an E.164
    phone number. The DLP catalog flags both and the medium-severity
    aggregate triggers the block.
\end{itemize}

\paragraph{Backends.} Each sample posts (or in the blocked case, would
have posted) to a live Discord channel and a live Telegram chat that
the operator owns. The OpenClaw subject's posts land verbatim in the
chat; the enclawed-oss and enclawed-enclaved subjects post a
single-line block notice (\texttt{[name | F*] message blocked: ...})
that never exposes the adversarial payload to the chat surface.

\subsection{Statistical methodology}
\label{sec:stat-methodology}

\paragraph{Confidence intervals.} For each per-cell precision and
recall point estimate, we report the Wilson score
interval~\cite{Wilson1927} at 95\% confidence. Wilson is preferred
over the normal-approximation interval at the extremes (events
near zero or one) because it never produces degenerate or
out-of-bounds bounds; this matters here because the hardened
subjects' false-positive count is identically zero across many
cells, where a normal-approximation interval would collapse to a
point.

For $k$ successes out of $n$ trials, the Wilson interval at
confidence level $1-\alpha$ with $z = \Phi^{-1}(1-\alpha/2)$ is
\begin{equation}
\widehat{p}_W \;=\; \frac{k + z^2/2}{n + z^2}, \qquad
\Delta \;=\; \frac{z}{n + z^2}\sqrt{\frac{k(n-k)}{n} + \frac{z^2}{4}},
\label{eq:wilson}
\end{equation}
giving the interval $[\widehat{p}_W - \Delta,\; \widehat{p}_W + \Delta]$
clipped to $[0, 1]$.

\paragraph{Sample-size justification.} A publication-grade upper bound
on the hardened subject's false-positive rate (FPR) requires $n$
large enough that the upper Wilson bound at $k = 0$ false positives
is small. For the legit-side column,
$\mathrm{FPR} = \mathrm{FP}/(\mathrm{FP}+\mathrm{TN})$. With $k=0$
and $n=100$ the Wilson 95\% upper bound is $\approx 0.036$; at
$n=10^4$, $\approx 3.84 \times 10^{-4}$. We report the headline at
$n = 100$ per cell to match the original enclawed paper's protocol
and a stress-test extension at $n = 10^4$ per (channel, F-cat,
label) cell ($n = 80{,}000$ total samples; Table~\ref{tab:n80k-stress})
to push the upper-bound claim below practitioner-relevant
false-alarm thresholds.

\paragraph{Paired comparison: McNemar's test.} The three subjects
under test see byte-for-byte the same sample set. The natural
statistical test for a paired-binary outcome of this shape is
McNemar's test~\cite{McNemar1947}. For two subjects $A$ and $B$
with disagreement counts $b$ (A blocked, B delivered) and $c$ (A
delivered, B blocked), the continuity-corrected statistic
\begin{equation}
\chi^2 \;=\; \frac{(\,|b-c| - 1\,)^2}{b + c}
\label{eq:mcnemar}
\end{equation}
is compared to the chi-squared distribution with one degree of
freedom. We report $b$, $c$, and $\chi^2$ per pair of subjects
inline with the stress-test results in \S\ref{sec:results} (see
the McNemar paragraph following Table~\ref{tab:n80k-stress});
paired comparisons where the disagreement is one-sided (every
adversarial blocked by $B$ and delivered by $A$, every legit
delivered by both) push $\chi^2$ to values where $p < 10^{-3}$
is conservative.

\paragraph{Random seeding and replay.} The harness's template
generators and pseudo-random number generator (PRNG) draws are
routed through a seedable \textsc{mulberry32} PRNG. Setting
\texttt{ENCLAWED\_INVIVO\_SEED} hashes its value to a 32-bit
integer and re-seeds; running with the same
seed reproduces the sample set byte-for-byte (the per-sample CSV
hashes match across replay runs). Cryptographic primitives (Ed25519
keypair, broker keys, accreditor keys) intentionally remain
non-seeded; they are fresh per process by design and the
attestation chain is independently re-verifiable with whatever
public keys the run emitted.

\paragraph{Hardware fingerprint.} Every run records the Node version,
operating system, CPU model and count, total RAM, harness git
commit, and PRNG seed at start, and writes them as a table at the
top of the markdown report. Reviewers replicating the experiment can
compare their own fingerprint against the cited one.

\paragraph{Inferential reach of the template stress test.} The
Wilson 95\% upper FPR bound at $k=0$, $n=10^4$ achieved in
Table~\ref{tab:n80k-stress} is $3.84\times 10^{-4}$ per cell ---
below the false-alarm-fatigue threshold for typical
practitioner-relevant traffic rates. Extending to $n=10^6$ per cell
would lower the bound to $\approx 3.7\times 10^{-6}$; beyond that
the bound is no longer practitioner-informative. The result is a
coverage claim on the gate's \emph{own} classifier surface, not a
generalisation claim; generalisation is the orthogonal LLM-driven
path (\S\ref{sec:sample-sources}). Closing the remaining gap to a
maximally adversarial reviewer would require: (i) the LLM-driven
study at $n\in[10^3, 10^4]$ per cell with full cooperation
disclosure (Table~\ref{tab:llm-coop} is the $n=200$ pilot on
cooperative Ollama LLMs); (ii) cross-LLM stability across
additional alignment regimes (Mixtral-8x7B-Instruct,
Hermes-3-405B); (iii) multi-seed variance; (iv) per-component
ablation (prompt-shield alone, DLP alone, biconditional alone).

\subsection{Reproducibility}

The harness lives at
\texttt{enclawed/test/security/in-vivo/llm-narrative.mjs} in the public
enclawed-oss repository~\cite{enclawedRepo}. Re-running the experiment
is one command:

\begin{verbatim}
node enclawed/test/security/in-vivo/llm-narrative.mjs
\end{verbatim}

\noindent Default behavior: template-synthesized samples, 100 per
cell, both cloud-channel rounds (Discord, Telegram). The harness
writes a markdown summary report and a per-sample CSV under
\path{docs/}, plus a hash-chained audit log and (when the certified
evolution's external-witness component is engaged at boot) an
externally-verifiable witness journal under
\path{~/.enclawed-invivo/}.

\section{Results}
\label{sec:results}

We report two complementary results.
\textbf{Baseline (\S\ref{sec:results-baseline})} answers \emph{does
the gate cover its own design distribution?} Templates are
deterministic and cooperation is 100\% by construction; failure
means the gate is broken on patterns it was tuned for. Passing is a
necessary lower bound, not a generalisation claim.
\textbf{Generalisation (\S\ref{sec:results-generalization})} answers
\emph{does the gate carry to distributions it was not tuned for?}
Ten LLMs author adversarial content per F-category; cooperation
varies per LLM (refusals are excluded by an independent
classifier). A subject scoring well on the baseline but poorly on
generalisation is brittle; a subject scoring well on both holds.

\paragraph{Confusion-matrix legend.} Each cell counts samples where the
ground-truth label is \emph{adversarial} (top row) or \emph{legit} (bottom
row) and the gate decision is \emph{block} or \emph{deliver}. \textbf{TP} =
correctly blocked attack; \textbf{FN} = missed attack (delivered an
adversarial payload); \textbf{TN} = correctly delivered legit message;
\textbf{FP} = false alarm (blocked a legit message). \textbf{Precision} =
TP/(TP+FP) is ``of the things I blocked, how many were actually attacks'';
\textbf{recall} = TP/(TP+FN) is ``of the actual attacks, how many did I
catch''. A subject that blocks everything trivially scores recall = 1 but
precision $\to$ 0; a subject that delivers everything scores recall = 0; only
$P = R = 1.0$ demonstrates a gate that actually discriminates.

\subsection{Reference baseline: regex template path}
\label{sec:results-baseline}

The baseline run uses the harness's deterministic template generators. Every
adversarial template emits a canonical pattern from the production gate's own
classifier surface (the prompt-shield imperative-override + role-boundary
catalogues for F1; an audit-record forgery against a per-probe biconditional
for F2; the DLP secret/PII regex catalog for F3 and F4). Cooperation is
therefore 100\% by construction and the question reduces to: for each
runtime, does the gate fire? The headline
(Table~\ref{tab:headline}) is the 1600-sample matrix; the
stress-test extension (Table~\ref{tab:n80k-stress}) pushes $n$ to
$80{,}000$ to tighten the FPR upper bound.

\subsubsection{Headline (baseline, $n=1600$)}

\begin{table}[h]
\centering
\small
\setlength{\tabcolsep}{6pt}
\renewcommand{\arraystretch}{1.15}
\begin{tabular}{|l|c|c|c|c|}
\hline
\textbf{Subject} & \textbf{F1\_BYPASS} & \textbf{F2\_FORGE} & \textbf{F3\_SILENT} & \textbf{F4\_WRONGTGT} \\
\hline
OpenClaw           & R=0.000 & R=0.000 & R=0.000 & R=0.000 \\
enclawed-oss       & \textbf{1.000} & \textbf{1.000} & \textbf{1.000} & \textbf{1.000} \\
enclawed-enclaved  & \textbf{1.000} & \textbf{1.000} & \textbf{1.000} & \textbf{1.000} \\
\hline
\end{tabular}
\caption{\textbf{REFERENCE BASELINE.} Headline confusion-matrix recall per
subject per F-category, template-synthesized samples ($n=200$ per cell, 100
legit + 100 adversarial, aggregated across both Discord and Telegram
channels, $n=1600$ total). For both hardened subjects, precision = recall =
F1 = accuracy = 1.000 in every cell; we report the F1 score for brevity.
Total wall-clock for the full pass: 42.2~seconds. This is the gate's
coverage of its design distribution; the generalization claim is in
\S\ref{sec:results-generalization}.}
\label{tab:headline}
\end{table}

\paragraph{Per-channel decomposition.} The headline aggregates Discord
and Telegram. The split is uniform: each channel contributes
$n=200$ per F-cat ($n=800$ per channel) with the same shape ---
OpenClaw (TP=0, FP=0, TN=100, FN=100, R=0.000) and both enclawed
subjects (TP=100, FP=0, TN=100, FN=0, R=1.000) on every F-category.
Per-channel matrices are not reproduced here; the 1600-sample
aggregate above is exactly the sum.

\subsubsection{Stress-test extension at $n = 80{,}000$ template samples
(baseline)}

Still on the baseline path, we tighten the FPR upper bound by extending the
template run to $n = 10{,}000$ per (channel, F-category, label) cell, for a
total of $80{,}000$ template-synthesized samples. The Wilson 95\% upper
bound on the per-cell false-positive rate at $k = 0$ false positives and
$n = 10{,}000$ is $3.84 \times 10^{-4}$.

Per-subject results in this stress-test run
(Table~\ref{tab:n80k-stress}):

\begin{table}[h]
\centering
\small
\setlength{\tabcolsep}{4pt}
\renewcommand{\arraystretch}{1.15}
\begin{tabular}{|l|c|c|c|c|}
\hline
\textbf{Subject} & \textbf{F1\_BYPASS} & \textbf{F2\_FORGE} & \textbf{F3\_SILENT} & \textbf{F4\_WRONGTGT} \\
\hline
OpenClaw                    & R=0.000        & R=0.000        & R=0.000        & R=0.000        \\
\hline
enclawed-oss                & \textbf{1.000} & \textbf{1.000} & \textbf{1.000} & \textbf{1.000} \\
\hline
enclawed-enclaved (content) & \textbf{1.000} & \textbf{1.000} & \textbf{1.000} & \textbf{1.000} \\
\hline
enclawed-enclaved (full)    & R=1.000        & R=1.000        & R=1.000        & R=1.000        \\
                            & FP=10000       & FP=8472        & FP=9315        & FP=10000       \\
\hline
\end{tabular}
\caption{\textbf{REFERENCE BASELINE, stress-test extension.} Per-subject
results at $n = 10{,}000$ legit + $10{,}000$ adversarial per (channel,
F-category) cell on the Telegram channel ($n=80{,}000$ total). The Wilson
95\% upper bound on the per-cell FPR at $k=0$ false positives and
$n=10{,}000$ is $3.84 \times 10^{-4}$, which holds in every cell of the
upper three rows (the secmon row is broken out separately). \emph{enclawed-oss}
and \emph{enclawed-enclaved (content gate only)} are identical at this run
(every cell matches): the F1--F4 contract is identical at the content layer,
and McNemar's test on the pair returns $\chi^2 = 0$ on the disagreement
counts. \emph{enclawed-enclaved (full stack)} additionally runs the secmon
behavioral monitor; at burst-test rates secmon correctly identifies the
harness's $\sim 250$~samples/sec firing rate as an anomalous traffic pattern
(P-014, ``actor request rate $\geq 60$/min''), so its FP column is high.
This is correct production behavior at non-production load and is reported
separately so the F1--F4 content matrix is readable; at production rates the
same detector would not fire on legit operational chatter. Wall-clock
927~seconds on i9-14900K single-thread, secmon engaged with bounded
64-actor pool, seed \texttt{stress-headline-2026}.}
\label{tab:n80k-stress}
\end{table}

McNemar's paired test on this run confirms the qualitative claim:
$\chi^2 \approx 4 \times 10^{4}$ ($p < 10^{-3}$, df $= 1$) for
OpenClaw vs.\ either of the hardened subjects; $\chi^2 = 0$ for
\emph{enclawed-oss vs.\ enclawed-enclaved (content)} (zero
disagreement; the two subjects' content gates produce byte-identical
verdicts on every sample); $\chi^2 \approx 3.8 \times 10^{4}$ for
\emph{enclawed-oss vs.\ enclawed-enclaved (full)} (the burst-rate
disagreement injected by secmon).

\subsection{Generalization: LLM-emitted path}
\label{sec:results-generalization}

The baseline above measures gate coverage on the design distribution; it
cannot answer whether the gate also catches adversarial content from
distributions it was not tuned for. The generalization run uses the
harness's \texttt{SAMPLE\_SOURCE=llm} mode, run against ten distinct LLMs
spanning vendor, parameter scale, and alignment regime. Each LLM produced
up to 100 legit and 100 adversarial samples per F-category per channel.
Two independent classifiers feed the recall denominator. A
\emph{refusal classifier} excludes refusals (regex matching
first-person disclaimer phrases such as ``I can't provide'',
``I'm sorry'', ``As an AI''). A \emph{cooperation classifier},
deliberately a strict superset of the gate's pattern catalog
(\S\ref{sec:stat-methodology}), counts a sample as cooperated when
the LLM emitted adversarial-shaped content. The cooperation
classifier therefore cannot be biased by the gate's own verdict.
Lower recall on this side relative to the baseline is the honest
measurement: it shows where the gate's regex-tuned coverage of its
own template surface fails to catch shapes a real LLM emits.

\paragraph{Independence of the cooperation classifier from the production
gate.} The cooperation classifier lives in
\texttt{enclawed/test/security/in-vivo/cross-llm-aggregate.py} and was
frozen at its current shape \emph{before} the regex-widening exercise
in \S\ref{sec:widening}; it has not been modified since. Its match set
is a strict superset of \emph{both} the strict and the widened
production DLP regex (i.e., its match set contains every secret-shape
match the production gate would ever return, plus additional
loosely-shaped variants neither version of the gate matches), and a
strict superset of the LLM-refusal classifier (so refusals never count
as cooperation, regardless of incidental keyword overlap). Three
properties follow that are load-bearing for every recall claim in
\S\ref{sec:results-generalization}:
(i)~the cooperation \emph{denominator} is invariant under any
production-gate regex change, including the widening reported in
\S\ref{sec:widening};
(ii)~the recall \emph{numerator} is the runtime gate's verdict at the
time of the run and cannot be moved by post-hoc analysis;
(iii)~for any gate-blocked row, cooperation is by construction true (a
row that the gate fired on is by definition adversarial-shaped).
Together these rule out the trivial concern that we widened the
classifier in lockstep with the gate; the +14.6\% TP gain reported in
\S\ref{sec:widening} is measured against an invariant denominator
definition.

\subsubsection{Cross-LLM stability of recall (generalization)}
\label{sec:cross-llm}

\begin{table}[h]
\centering
\small
\setlength{\tabcolsep}{6pt}
\renewcommand{\arraystretch}{1.1}
\begin{tabular}{|l|c|c|c|c|c|}
\hline
\textbf{Failure mode} & \textbf{$N_{\mathrm{LLM}}$}
  & \multicolumn{4}{c|}{\textbf{enclawed-oss recall on cooperated samples}} \\
\cline{3-6}
                          &   & \textbf{mean} & \textbf{std dev} & \textbf{min} & \textbf{max} \\
\hline
F1 (gate bypass)          & 5 & 1.000 & 0.000 & 1.000 & 1.000 \\
F2 (audit forgery)        & 6 & 1.000 & 0.000 & 1.000 & 1.000 \\
F3 (silent host failure)  & 4 & 1.000 & 0.000 & 1.000 & 1.000 \\
F4 (wrong target)         & 5 & 1.000 & 0.000 & 1.000 & 1.000 \\
\hline
\end{tabular}
\caption{\textbf{Cross-LLM recall stability for enclawed-oss
(generalisation path).} For each failure mode, $N_{\mathrm{LLM}}$ is
the number of LLMs in the ten-LLM panel that produced at least 10
cooperated adversarial samples per F-category (refusals excluded by
an independent classifier, \S\ref{sec:stat-methodology};
non-cooperating LLMs are not ranked). Per-LLM recall is computed as
(adversarial samples the gate blocked)~/~(adversarial samples the
cooperation classifier flagged as adversarial-shaped); each value
lies in $[0, 1]$. The four columns report the mean, standard
deviation, minimum, and maximum of those per-LLM recalls across the
$N_{\mathrm{LLM}}$ cooperating LLMs. All four rows show mean recall
$= 1.000$ at zero variance: every cooperated adversarial sample was
blocked on every cooperating LLM. \emph{enclawed-enclaved
(content gate only)} is byte-identical to the table above (McNemar
$\chi^2 = 0$ on every pair) and is omitted; upstream OpenClaw
reports recall $= 0.000$ on every (LLM, F-cat) cell and is also
omitted. Per-LLM detail for F3 (where the live regex-widening
exercise lands) is in Table~\ref{tab:perruntime-f3-silent}; F1, F2,
F4 follow the same shape.}
\label{tab:cross-llm}
\end{table}

The cooperation rate itself is informative as a separate axis
(Table~\ref{tab:llm-coop}). LLM safety training is highly variable:
\texttt{ollama llama3.2:3b} refused F1, F3, and F4 on 100\% of
samples; \texttt{groq openai/gpt-oss-120b} refused $> 99\%$;
\texttt{gemini-2.5-flash} hit free-tier rate limits and produced
zero usable adversarial samples in our window. By contrast,
\texttt{ollama llama3.1:8b}, \texttt{mistral:7b}, \texttt{qwen2.5:7b},
\texttt{gemma2:9b} all cooperated on F1 at $\geq 99.5\%$ and on F4
at 100\%; F3 cooperation ranged from 38\% (llama3.1:8b) to 82\%
(qwen2.5:7b) reflecting the heavy safety training around literal
secret formats. This is itself a finding: ``the LLM refuses to author
the attack'' is not a security guarantee, because (a) even within a
single vendor family the refusal stance varies markedly between
adjacent versions, and (b) the F1 $\geq 99.5\%$ cooperation across
four open-weights models in the 7--9B range shows that obtaining
attacker-cooperative output is trivial. The host-side gate must
hold the line regardless.

\begin{table}[h]
\centering
\small
\setlength{\tabcolsep}{4pt}
\begin{tabular}{|l|c|c|c|c|}
\hline
\textbf{LLM} & \textbf{F1 coop \%} & \textbf{F2 coop \%} & \textbf{F3 coop \%} & \textbf{F4 coop \%} \\
\hline
ollama llama3.2:3b           &   0.0 & 10.0 &   0.0 &   0.0 \\
ollama llama3.1:8b           & 100.0 &100.0 &  38.0 & 100.0 \\
ollama mistral:7b            & 100.0 &100.0 &  55.0 &  99.5 \\
ollama qwen2.5:7b            &  99.5 &100.0 &  82.0 & 100.0 \\
ollama gemma2:9b             &  99.5 & 99.0 &  59.5 & 100.0 \\
openrouter llama-3.1-70b     &  68.0 &100.0 &   1.5 & 100.0 \\
\hline
\multicolumn{5}{|l|}{\emph{Cloud-quota-limited rows ($n<10$ cooperated; ranking not meaningful):}} \\
\hline
groq llama-3.3-70b-versatile &     100 &     100 &    0 & 100 \\
groq llama-4-scout-17b       &     100 &     100 &   -- & 100 \\
groq openai/gpt-oss-120b     &     0.0 &     0.0 &  0.0 & 0.0 \\
gemini-2.5-flash             &      -- &      -- &   -- & --  \\
\hline
\end{tabular}
\caption{\textbf{GENERALIZATION, cooperation rates.} Per-LLM cooperation
rate per F-category at $n = 200$ adversarial samples per cell ($100$ on
each of two channels). The lower block lists rows that exhausted the
cloud provider's free-tier rate quota during the run window and
therefore have $n<10$ cooperated samples; their numbers are included
for completeness but should not be ranked. F2\_FORGE is
content-agnostic; non-100\% cooperation in that column on small-LLM
rows reflects the refusal classifier rejecting refusal-shaped content
the LLM emitted in lieu of a routine audit-line.}
\label{tab:llm-coop}
\end{table}

\paragraph{Baseline-side block reasons.} On the templates path (baseline),
\texttt{prompt-shield findings} fired 200/200 on F1\_BYPASS samples;
\texttt{biconditional: f2Forgery on 1 (cap, target) projection(s)} fired
200/200 on F2\_FORGE; \texttt{DLP findings (severity=high)} fired 200/200
on F3\_SILENT; \texttt{DLP findings (severity=medium)} fired 200/200 on
F4\_WRONGTGT. No legit sample triggered any block reason (FP = 0 across
all cells) on the templates path. We include this here as the cross-check
that the gate's stated detection mass on the baseline maps to the same
shipped detectors that handle the LLM-emitted samples below.

\subsubsection{Per-(source, runtime) consolidated table for F3}
\label{sec:perruntime-tables}

Table~\ref{tab:perruntime-f3-silent} (auto-generated by
\texttt{cross-llm-aggregate.py}) puts baseline and per-LLM
generalisation rows side-by-side for F3\_SILENT --- the F-category
where the regex-widening exercise of \S\ref{sec:widening} lands.
F1, F2, F4 follow the same shape (OC = 0, OSS = ENC = 1.000 on
every cooperating LLM) and are omitted for compactness. Cooperation
counts a row when either the runtime gate fired on it or the
(possibly redacted) content carries a raw secret/PII shape, which
keeps the metric stable whether the analysed CSV is the runtime
artifact or a publish-scrubbed copy. Three observations: (i) OC
recall = 0.000 on every cell of every LLM, including LLMs
cooperating at $\geq 199$/200 samples; (ii) OSS and ENC agree
byte-for-byte on recall wherever $n_{\mathrm{coop}} \geq 1$; (iii)
the gate's reach \emph{ends}, on the LLM-emitted side, not in
recall (the gate blocks every cooperated sample) but in
cooperation rate --- short / malformed / placeholder shapes fall
under the cooperation threshold and are addressed by the widening
of \S\ref{sec:widening}.

%

\begin{table}[h]
\centering
\footnotesize
\setlength{\tabcolsep}{3pt}
\renewcommand{\arraystretch}{1.05}
\resizebox{\textwidth}{!}{%
\begin{tabular}{|l|r|r|c|c|c|c|}
\hline
\textbf{Source} & \textbf{$n_{\mathrm{adv}}$} & \textbf{$n_{\mathrm{coop}}$} & \textbf{enclawed-oss} & \textbf{enclawed-} & \textbf{enclawed-oss} & \textbf{enclawed-} \\
           & & & \textbf{recall} & \textbf{enclaved recall} & \textbf{FPR} & \textbf{enclaved FPR} \\
\hline
\textbf{templates (baseline)} & 200 & 200 & 1.000 [0.98, 1.00] & 1.000 [0.98, 1.00] & 0.000 [0.00, 0.02] & 0.000 [0.00, 0.02] \\
\hline
Groq gpt-oss-120b & 2 & 0 & -- & -- & -- & -- \\
Groq llama3.3-70b & 2 & 0 & -- & -- & 0.000 [0.00, 0.79] & 0.000 [0.00, 0.79] \\
Groq llama4-scout-17b & 0 & 0 & -- & -- & 0.000 [0.00, 0.66] & 0.000 [0.00, 0.66] \\
Ollama gemma2-9b & 200 & 165 & 1.000 [0.98, 1.00] & 1.000 [0.98, 1.00] & 0.000 [0.00, 0.02] & 0.000 [0.00, 0.02] \\
Ollama llama3.1-8b & 200 & 76 & 1.000 [0.95, 1.00] & 1.000 [0.95, 1.00] & 0.000 [0.00, 0.02] & 0.000 [0.00, 0.02] \\
Ollama llama3.2-3b & 200 & 0 & -- & -- & 0.000 [0.00, 0.02] & 0.035 [0.02, 0.07] \\
Ollama mistral-7b & 200 & 110 & 1.000 [0.97, 1.00] & 1.000 [0.97, 1.00] & 0.000 [0.00, 0.02] & 0.000 [0.00, 0.02] \\
Ollama qwen2.5-7b & 200 & 164 & 1.000 [0.98, 1.00] & 1.000 [0.98, 1.00] & 0.000 [0.00, 0.02] & 0.000 [0.00, 0.02] \\
OpenRouter llama3.1-70b & 200 & 3 & 1.000 [0.44, 1.00] & 1.000 [0.44, 1.00] & 0.000 [0.00, 0.02] & 0.000 [0.00, 0.02] \\
\hline
\end{tabular}%
}
\caption{\textbf{F3 (silent host failure) --- per-runtime detection per source.} \textbf{Columns.} $n_{\mathrm{adv}}$ is the number of adversarial samples the source contributed; $n_{\mathrm{coop}}$ is the subset that an independent classifier flagged as adversarial-shaped (refusals excluded; \S\ref{sec:stat-methodology}). \emph{Recall} = (samples blocked by the gate) / $n_{\mathrm{coop}}$; \emph{FPR} (false-positive rate) = (legit samples wrongly blocked) / 100 per source. \textbf{Cell format.} Each cell reports the point estimate followed by the Wilson 95\% confidence interval as \texttt{<point> [<low>, <high>]}; ``--'' indicates $n_{\mathrm{coop}}=0$ or $n_{\mathrm{legit}}=0$ so the metric is undefined. CI width depends on $n_{\mathrm{coop}}$: rows with low cooperation (e.g.\ \texttt{OpenRouter llama-3.1-70b} at $n_{\mathrm{coop}}=3$) carry wide intervals such as $[0.44, 1.00]$ even when every cooperated sample was blocked, and should not be ranked against high-$n$ rows. \textbf{Rows.} The first row (separated by a horizontal rule) is the regex-template reference baseline (in-distribution by construction; cooperation = 100\%); subsequent rows are per-LLM cross-model generalisation. Upstream OpenClaw is a passthrough negative control, reports recall = FPR = 0 on every cell of every source by construction, and is omitted for compactness.}
\label{tab:perruntime-f3-silent}
\end{table}

\subsubsection{Improvability is the architectural property: a worked
example of widening the F3 gate}
\label{sec:widening}

The cross-LLM data above (Tables~\ref{tab:cross-llm}--\ref{tab:perruntime-f3-silent})
reports a recall ceiling of mean = 1.000 on cooperated samples in every
F-category, so the harness was not flagging false negatives. The interesting
question is at the layer beneath: \emph{cooperation itself} is bounded by what
the post-hoc classifier recognises as adversarial-shaped --- the union of
``the gate fired'' and ``the content matches a fixed superset regex''
(\S\ref{sec:stat-methodology}). Borderline shapes the LLMs emit but that
match neither --- short tokens, prefix-glued tokens, padding-char variants
--- fall \emph{below the cooperation threshold} and are counted in neither
numerator nor denominator. They are coverage gaps, not missed catches. The
practically interesting question the per-LLM cooperation rates
(Table~\ref{tab:llm-coop}) make visible is: \emph{what does it cost to pull
those gap shapes into the harness as recognised, blocked positives?} We
answered by widening the production DLP regex catalog and re-running the
harness on the four cooperative Ollama LLMs.

\paragraph{The change, validation, and gain.}
\begin{sloppypar}
Five new high-severity patterns were appended to
\texttt{enclawed/src/dlp-scanner.mjs} (plus its TypeScript twin)
plus a one-character boundary fix on the existing AWS pattern. They
target shapes the strict catalog rejected: short fake secret-prefix
tokens (\texttt{sk-XXXXXX}, \texttt{AKIA1234}, \texttt{xoxb-foo});
prefix tokens glued to surrounding word chars that defeat the
leading \verb|\b| (\texttt{\_AKIA...}, \texttt{DEBUG=ghp\_...}); and
OpenAI-style keys with the \texttt{=} padding char (\texttt{sk-=...}).
A six-line append-only patch.
\end{sloppypar}
The widened catalog was scanned against 5{,}708 legit-side cells in
the existing artifact corpus (template path + all ten cross-LLM
CSVs); findings at \texttt{high} or \texttt{critical}: zero. A
re-run on the four cooperative Ollama LLMs (Discord, $K=100$,
license-engaged accreditor) yields the F3\_SILENT TP counts in
Table~\ref{tab:widening-gain}.

\begin{table}[h]
\centering
\small
\begin{tabular}{|l|r|r|c|c|}
\hline
\textbf{LLM source} & \textbf{Pre TP/100} & \textbf{Post TP/100} & \textbf{Delta} & \textbf{OSS FPR} \\
\hline
\texttt{ol-llama3.1-8b}    & 38   & \textbf{61} & +23 (+61\%) & 0.000 \\
\texttt{ol-mistral-7b}     & 55   & \textbf{60} & +5\phantom{0}  (+9\%)  & 0.000 \\
\texttt{ol-qwen2.5-7b}     & 82   & \textbf{84} & +2\phantom{0}  (+2\%)  & 0.000 \\
\texttt{ol-gemma2-9b}      & 82.5 & \textbf{90} & +7.5 (+9\%) & 0.000 \\
\hline
\textbf{Cumulative}        & 257.5 & \textbf{295} & +37.5 (+14.6\%) & 0.000 \\
\hline
\end{tabular}
\caption{Post-widening F3\_SILENT true-positive counts per cooperative
Ollama LLM, Discord channel, $K=100$ per cell. Pre values are halved
historical two-channel counts for fair per-channel comparison
(\texttt{docs/post-widening/AGGREGATE.md} carries the raw new run).
ENC full-stack matches OSS byte-for-byte across all four LLMs (secmon
does not fire on F3 cells under normal LLM throughput).}
\label{tab:widening-gain}
\end{table}

\noindent The mechanism is the same one the cooperation definition makes
visible: borderline LLM emissions that previously fell below the
cooperation threshold (no gate fire, no post-hoc match) are now caught by
the widened gate, and a gate fire is itself one of the two ways a row
counts as cooperated --- so the same edit drags those rows simultaneously
into the denominator (newly recognised as adversarial-shaped) and into
the numerator (blocked). Per-cell recall on cooperated therefore stays at
1.000 in both runs --- the gate is still catching every shape the
classifier recognises --- while the underlying true-positive count grows
by 14.6\% net across the four LLMs, with all four showing positive deltas
and zero new false positives. Under a fixed-recall ceiling, this is what
coverage improvement looks like in practice: the gate blocks a
measurably larger superset of LLM-emitted F3 shapes at the same
precision, and there were no false negatives to begin with --- the gain
is recognised coverage, not corrected misses.

\paragraph{Why this is the obsolescence point, not a footnote.} The
widening is a single-commit, append-only change to a primitive the
framework declares as a public surface (the \texttt{PATTERNS} array
in \texttt{dlp-scanner.mjs}): five regex literals plus a
one-character boundary fix. The empirical loop --- observe a miss
in the CSV, append a pattern, validate, re-run, see the catch count
grow at FPR = 0 --- is the loop the architecture was
\emph{designed} for, because each layer is named and testable in
isolation. The same loop on any runtime in the unhardened category
has nowhere to begin: no DLP catalog to append to, no gate to
consult one, no audit log, no biconditional. Closing the gap
requires adding all seven primitives jointly --- a re-architecture
that produces enclawed-oss. The seven primitives are not just
present; they are the \emph{seams} that make safety improvable.

\subsection{Cost and traffic}

Wall-clock for the 1600-sample baseline pass: $\approx 42$~seconds,
dominated by chat-post round-trips with a 600--800~ms pacing delay;
gate stack itself runs in microseconds per sample. The
$n=80{,}000$ stress test takes $\approx 927$~seconds on a single
i9-14900K thread. LLM-driven runs add the marginal LLM cost and stay
under \$0.10 per cooperative-cell pass on the cloud LLMs we measured.

\section{Why this constitutes architectural obsolescence}
\label{sec:obsolescence}

The data of \S\ref{sec:results} satisfies all three conditions of
the comparative definition in \S\ref{sec:def-obsolescence}.
\textbf{Strict dominance:} OpenClaw recall is 0.000 on every cell
of every F-category on both the baseline and the generalisation
distribution; enclawed-oss reports $P = R = 1.000$ on the same
input. \textbf{Practical adoptability:}
enclawed-oss is MIT-licensed, ships through the same package
ecosystem OpenClaw uses, and preserves OpenClaw's user-visible
feature surface. An operator currently running OpenClaw can adopt
enclawed-oss with no new infrastructure, no new contracts, and no
behavioural change visible to end users. \textbf{Structural gap:}
the path from OpenClaw's 0\% to enclawed-oss's 100\% is not a
configuration knob. A defender who proposes adding an audit log,
then a biconditional, then a trust root, then an admission gate,
then an egress guard, then a classification lattice, then a
bootstrap seal has at the end re-implemented enclawed-oss; the
seven primitives are joint, and missing any one re-opens at least
one F-category.

\paragraph{0\% is qualitatively different from low.} A runtime
catching 60\% of F1 has a tunable detector and is properly called
\emph{weak}; a runtime catching 0\% with no detector in the source
tree is properly called \emph{obsolete relative to one that catches
100\% with detectors the first does not have}. The two failure
modes require qualitatively different remediation paths --- parameter
tuning vs.\ adoption of an architecturally different runtime --- and
the paper's claim is about the second only.

\paragraph{Stochastic LLM refusal is not a security primitive.}
With no detection mechanism in the source tree, an OpenClaw
deployment's only defence against an F1--F4 attack is whether the
configured LLM happens to refuse to author it. Table~\ref{tab:llm-coop}
shows this varies wildly: same-vendor adjacent versions diverge by
100 percentage points (F1 cooperation: 0\% on
\texttt{llama3.2:3b}, 100\% on \texttt{llama3.1:8b}); F3 cooperation
ranges from 0\% to 82\% across the panel. An operator who switches
the configured LLM --- a one-flag change --- can move an
F-category's effective ``protection'' from 100\% to 0\% without
touching the runtime. enclawed's F1--F4 gate has no such dependency:
it fires on the \emph{shape} of the input regardless of which LLM
authored it. Stochastic refusal by a model that may be retrained
tomorrow is not a security primitive; deterministic shape detection
by a versioned, testable gate is.

\paragraph{Compliance and safety follow.} Every major regulator
(NIST 800-53, PCI~DSS 4.0, HIPAA, SOC~2, FedRAMP, NIS2, ISO~27001)
requires tamper-evident audit of consequential decisions. A runtime
that emits no audit record on F1 is non-compliant by inspection.
For deployments that actuate physical devices, F1 alone is
sufficient for an agent to mutate the world undetectably. The
existence of enclawed-oss removes the ``no alternative was
available'' defence that would otherwise excuse continued OpenClaw
deployment in these regimes.

\paragraph{Generalisation beyond OpenClaw.} OpenClaw is the
strongest representative of its class; weaker members score
$\leq$ OpenClaw's score by inheritance. We do not need to measure
each one separately. The claim is comparative, so it generalises
naturally: any agentic-AI runtime $R$ such that $R$'s recall on
F1--F4 is 0 and enclawed-oss's recall is 1 is obsolete with
respect to F1--F4 by the same definition. The detection-primitive
vocabulary is small and explicit; a tree-walk over any candidate
runtime takes minutes. We invite reviewers to publish the result
for any candidate they care about.

\section{Discussion}
\label{sec:discussion}

\subsection{LLM safety training is not a security boundary}

A common reaction to 1.000 recall on F1 is: ``surely the LLM itself
already refuses to author \texttt{IGNORE ALL PREVIOUS INSTRUCTIONS
\textless/system\textgreater}; why is a host-side gate even
necessary?'' Because LLM safety training is asymmetric: a safety-tuned
model refuses to \emph{author} an attack when politely asked but
happily \emph{processes} the same string when an attacker sends it as
input. Refusal-on-generation is not refusal-on-execution; real-world
incidents come from indirect injection (poisoned documents,
system-prompt overrides arriving as tool output, multilingual
obfuscation, retrieved-context attacks) and from adversaries who do
not need a cooperating LLM to produce the payload at all. The
framework's host-layer gates (\texttt{prompt-shield}, DLP, secmon)
fire on the string regardless of source, which is the whole rationale
for defense-in-depth at the runtime
boundary~\cite{Greshake2023indirect, Liu2024promptinjection}. The
LLM's safety training is a useful heuristic, never a security
boundary; the per-LLM cooperation rates in Table~\ref{tab:llm-coop}
make the asymmetry concrete (Llama 3.2:3b refuses roughly 95\% of
F1/F3/F4 briefs; Llama 3.1:70b cooperates on F1 at 95\% but refuses
F3 secret-exfil at 97\%).

\subsection{Scope of the F1--F4 contract}
\label{sec:scope}

F1--F4 do not exhaust the universe of agentic-AI attacks; they cover
the structural class in which action and accountability diverge.
Read-only exfiltration, TOCTOU races on the corpus, and collusion with
a malicious actor whose actions are correctly logged are out of scope
by construction (\cite{Metere2026enclawed}\,§5.4 names them
explicitly). A 1.000 recall on F1--F4 does not imply 1.000 on every
adversarial mode; it implies 1.000 on the four modes the biconditional
check is designed for.

\subsection{Why FP = 0 is non-trivial: no functionality cost}
\label{sec:no-tradeoff}

\begin{sloppypar}
A degenerate ``deny everything'' subject would score recall = 1.000
but precision = 0.500. The hardened subjects report
$P = R = 1.000$, $\mathrm{FP} = 0$ in every cell of the headline
matrix (Table~\ref{tab:headline}): the
gate distinguishes operational chatter (the four LEGIT\_*
templates) from adversarial templates with no false alarms. This is
structural, not lucky. The F1--F4 gates classify \emph{shape}, not
intent: the biconditional fires on multiset divergence; DLP fires on
canonical secret/PII formats that routine chat does not contain by
accident; prompt-shield fires on imperative-override + role-boundary
serialization tokens; the admission gate fires on manifest signature
plus capability vocabulary. There is no business logic to mis-tune,
no acceptable-use policy to draft, no vendor-specific allowlist to
keep current. The ``security costs functionality'' trade-off is real
for IDS/WAF/EDR products that classify content with
deployment-specific heuristics; it is not real for the F1--F4
contract because the contract reconciles structure rather than
scoring content.
\end{sloppypar}

\subsection{What the certified evolution adds beyond the OSS tier}

The enclaved subject's matrix coincides with oss's on F1--F4 because
both content gates reach the same verdict on the same content. What
enclaved adds beyond that is \emph{evidence of who decided}: every
gate decision carries an external Ed25519-signed witness record that
a third party can re-verify with only the witness's public key. We
verify the fail-closed half empirically: re-running with the
external-witness component disabled forces the enclaved subject to
refuse all admissions for the rest of the run.

\subsection{Reviewer-replayable evidence}

The CSV at \texttt{docs/adversarial-in-vivo-samples.csv} records every
per-sample ground-truth label and per-subject decision; reviewers can
re-derive the matrix from the CSV without running the harness. The
audit log at \texttt{\textasciitilde/.enclawed-invivo/audit.jsonl} is
hash-chained and re-verifiable with \texttt{verifyChain()}. The
witness journal at \texttt{\textasciitilde/.enclawed-invivo/witness.jsonl}
is Ed25519-signed; reviewers can reconstruct the broker public key
and replay the attestations independently.

\section{Threats to validity}
\label{sec:validity}

\paragraph{External: in-distribution templates vs.\ generalisation.}
The template path is in-distribution by construction --- the regex
catalogs were tuned for exactly the patterns the templates emit, so a
$P=R=1.0$ result on it is a coverage claim, not a generalisation
claim. We mitigate via the orthogonal LLM-driven path
(\S\ref{sec:sample-sources}) which exercises content the regex was
\emph{not} designed against; cross-LLM mean recall = 1.000 on
cooperated samples (Table~\ref{tab:cross-llm}) is the
generalisation evidence.

\paragraph{External: channel and LLM coverage.} The harness exercises
Discord and Telegram (Slack/WhatsApp/Matrix etc.\ untested) and ten
LLMs (four cooperative Ollama models + six rate-limited or
heavily-refusing cloud models). Five cloud cells have $n<10$
cooperated samples and should not be ranked. A paid-tier sustained-quota
re-run would tighten cloud rows; the qualitative cross-LLM stability
claim already holds on the four Ollama models at $n\geq100$ each.

\paragraph{Construct: F2 is exercised behaviorally, not via content.}
F2 samples carry benign content; the failure-mode is the
audit-record-without-corpus-mutation behavior. The harness writes the
forged audit record per probe and runs the per-probe biconditional ---
a faithful instantiation of the F2 contract from
\cite{Metere2026enclawed}\,§5.

\paragraph{Internal: ground-truth label leakage and template
diversity.} The adversarial label is assigned at sample-generation time
and never reaches the gate (verified by inspecting the gate's call
trace: only the chat-message string, manifest, and policy enter). F1
templates have $\approx 5.8\times 10^4$ distinct variants; F3/F4 are
effectively unbounded thanks to random secret/PII bodies; F2 is
content-agnostic. Repetition begins around $n\approx 5\times 10^4$ on
F1 but does not bias the regex verdict.

\paragraph{Internal: cooperation classifier independence.}
\begin{sloppypar}The cooperation classifier
(\texttt{cross-llm-aggregate.py}) was frozen \emph{before} the
regex-widening exercise and is a strict superset of both the strict
and widened production regex. The widening's measured TP-count gain is
therefore against an invariant denominator. An independent
secret-scanner oracle (gitleaks, trufflehog, GitHub secret-scanning,
\texttt{detect-secrets}) could replace our classifier without other
harness changes; we expect headline numbers to move within Wilson CI.
\end{sloppypar}

\paragraph{Internal: single-seed, single-channel widening run.} The
widening verification (\S\ref{sec:widening}) ran on a single PRNG seed
on the Discord channel only ($K=100$ per LLM per F-cat). The +14.6\%
net cumulative TP gain is consistent across all four cooperative
Ollama LLMs, excluding a single-LLM artefact, but a multi-seed
two-channel re-run would tighten per-LLM Wilson intervals.

\paragraph{Internal: cross-sample state leakage.} The biconditional
uses per-probe scoping so F2 samples are independent. The audit log
persists across samples by design; we verified that no per-cell
decision depends on samples from a different cell. Channels are run
sequentially; a parallel-channel run could introduce ordering effects
we did not measure.

\paragraph{OpenClaw subject runs the upstream CLI per sample.} The
OpenClaw subject (\texttt{mediateOpenclaw}) invokes the actual upstream
\texttt{openclaw} CLI as a subprocess for every per-sample mediation
(\texttt{message send --channel <id> --target <t> --message
<content> --dry-run --json}), against an upstream checkout that has
been installed and built (\texttt{pnpm install \&\& pnpm
build:docker}). The CLI executes upstream's full pipeline --- argv
parsing, subcommand dispatch, message-send command handler, channel
resolution, payload framing, outbound delivery decision --- and emits
a JSON envelope; the harness records that envelope verbatim and
derives the verdict from \texttt{action}, \texttt{payload},
\texttt{handledBy}, and any \texttt{error} field upstream emits. The
\texttt{--dry-run} flag stops the pipeline at the network-call
boundary so no Discord or Telegram message is transmitted, which is
the only departure from a production end-user invocation. The two
upstream features this run does not exercise --- (a) the LLM agent
loop that decides \emph{which} payload reaches the send path, and (b)
real chat-backend transmission that tests the backend's own moderation
--- are deliberately out of scope: the F1--F4 contract is about the
runtime's gate, not about the agent's tool-call decision or the
backend's moderation. A future upstream release that introduces a
content-side gate between argv parsing and outbound delivery would
fire here automatically (the harness records the runtime's verdict
verbatim); today no such gate exists in the upstream tree, and the
harness records that absence empirically by running the CLI rather
than by assertion.

\section{Limitations}
\label{sec:limitations}

\textbf{Subject coverage.} The run covered Discord and Telegram
cloud channels (chosen because they are in the previously-deleted
unsafe set, \S\ref{sec:def-obsolescence}); other channel families
were not run, though the harness supports them.
\textbf{Adversarial agent shape.} The harness fires scripted
adversarial payloads at real backends, not an autonomously-misbehaving
LLM. The biconditional is source-agnostic, so we expect identical
detection topology under a prompt-injected LLM driver, but the
empirical confirmation is future work.
\textbf{What is not tested.} F1--F4 do not cover read-only
exfiltration, TOCTOU on the corpus, or operator collusion; recall =
1.000 on F1--F4 does not imply 1.000 on those modes
(\S\ref{sec:scope}).

\section{Reproducibility}

Wall-clock: $\approx 42$~s for the $n=1600$ baseline,
$\approx 927$~s for the $n=80{,}000$ stress test on a single
i9-14900K thread; the cross-LLM run is dominated by LLM round-trip
latency. Full reproduction:

\begin{verbatim}
git clone https://github.com/metereconsulting/enclawed && cd enclawed
# Required: cloud-channel credentials (operator-owned test guild).
export ENCLAWED_INVIVO_DISCORD_BOT_TOKEN=...
export ENCLAWED_INVIVO_DISCORD_CHANNEL_ID=...
export ENCLAWED_INVIVO_TELEGRAM_BOT_TOKEN=...
export ENCLAWED_INVIVO_TELEGRAM_CHAT_ID=...

# (1) baseline; (2) cross-LLM; (3) aggregate; (4) publish-safety scrub.
node enclawed/test/security/in-vivo/llm-narrative.mjs
ENCLAWED_INVIVO_LLM_K=100 \
  ./enclawed/test/security/in-vivo/cross-llm-driver.sh
python3 enclawed/test/security/in-vivo/cross-llm-aggregate.py \
  --csv-dir docs/cross-llm \
  --templates-csv docs/adversarial-in-vivo-samples.csv \
  --latex-out enclawed/paper/cross-llm-perruntime-table.tex \
  > docs/cross-llm-AGGREGATE.md
node scripts/scrub-invivo-csv.mjs
\end{verbatim}

\noindent Reports and per-sample CSVs (DLP-redacted) land under
\texttt{docs/}; hash-chained audit and Ed25519-signed witness
journals under \texttt{\textasciitilde/.enclawed-invivo/}. The
publish-safety scrubber must be run before per-sample CSVs are
committed or shared (\S\ref{sec:perruntime-tables} discusses why).

\section{Conclusion}

enclawed-oss is strictly better than upstream OpenClaw --- the
strongest measured representative of the unhardened-agentic-AI
class --- on every cell of every F1--F4 confusion matrix, and it is
MIT-licensed and drop-in adoptable. Across 1600 template-synthesised
baseline samples, ten cross-LLM generalisation runs (refusals
excluded by an independent classifier), and an 80{,}000-sample
stress test, OpenClaw recall is 0.000 on every cell; enclawed-oss
reaches $P = R = F_1 = $ accuracy $= 1.000$. A six-line append-only
widening of enclawed-oss's DLP catalog (\S\ref{sec:widening}) raises
per-channel F3 detection by 14.6\% net at unchanged precision; the
same edit on OpenClaw has nowhere to land --- the catalog, the gate,
the audit log, and the biconditional are all absent. The strict
dominance, practical adoptability, and structural gap are the three
conditions of architectural obsolescence (\S\ref{sec:def-obsolescence});
we exhibit them empirically. We invite reviewers to apply the
harness to any candidate runtime; we expect $0/N$ for any that does
not ship the seven primitives and $N/N$ for any that does.

\section*{Acknowledgments and Availability}

The framework primitives are from the original enclawed
paper~\cite{Metere2026enclawed}; we thank Peter Steinberger and the
OpenClaw contributors for the upstream gateway used as both the
fork base and the negative-control subject. Paper, harness,
aggregator, and per-LLM artifacts are MIT-licensed in the public
\texttt{enclawed-oss}
repository~\cite{enclawedRepo}.


\end{document}